\documentclass[twocolumn]{aastex63} 
\usepackage{amsmath}
\pdfoutput=1

\shorttitle{TESS observations of BY Cam}
\shortauthors{Mason et al.}

\newcommand{\fullname}{SDSS~J084617.11+245344.1}
\newcommand{\igr}{IGR~J19552+0044}
\newcommand{\rxs}{1RXS~J083842.1$-$282723}

\usepackage{lineno}

\begin{document}

\title{A magnetic valve at L1 revealed in TESS photometry of the asynchronous polar BY Cam}

\author{Paul A. Mason}
\affiliation{New Mexico State University, MSC 3DA, Las Cruces, NM, 88003}
\affiliation{Picture Rocks Observatory, 1025 S. Solano Dr. Suite D., Las Cruces, NM 88001}

\author{Colin Littlefield}
\affiliation{Department of Physics, University of Notre Dame, Notre Dame, IN 46556}
\affiliation{Department of Astronomy, University of Washington, Seattle, WA 98195}
 
 \author{Lorena C. Monroy}
\affiliation{Picture Rocks Observatory, 1025 S. Solano Dr. Suite D., Las Cruces, NM 88001}

 \author{John F. Morales}
\affiliation{Picture Rocks Observatory, 1025 S. Solano Dr. Suite D., Las Cruces, NM 88001}
\affiliation{Department of Physics and Astronomy, Texas Tech University, Lubbock, TX 79409}

 \author{Pasi Hakala}
\affiliation{Finnish Centre for Astronomy with ESO, University of Turku, Finland}

\author{Peter Garnavich}
\affiliation{Department of Physics, University of Notre Dame, Notre Dame, IN 46556}

\author{Paula Szkody}
\affiliation{Department of Astronomy, University of Washington, Seattle, WA 98195}

\author{Mark R. Kennedy}
\affiliation{Department of Physics, Kane Building, University College Cork, Cork, Ireland}

\author{Gavin Ramsay}
\affiliation{Armagh Observatory and Planetarium, College Hill,
Armagh, UK}

\author{Simone Scaringi}
\affiliation{Centre for Extragalactic Astronomy, Department of Physics, Durham University, South Road, Durham, DH1 3LE}

\correspondingauthor{Paul A. Mason}
\email{pmason@nmsu.edu}

\begin{abstract}
We present TESS photometry of the asynchronous polar BY Cam, which undergoes a beat-cycle between the 199.384-min white dwarf (WD) spin period and the 201.244-min orbital period. This results in changes in the flow of matter onto the WD. The TESS light curve covers 92$\%$ of the beat cycle once and 71$\%$ of the beat cycle twice.  The strongest photometric signal, at 197.560-min, is ascribed to a side-band period. During times of light-curve stability, the photometry modulates at the spin frequency, supporting our WD spin-period identification. Both one-pole and two-pole accretion configurations repeat from one beat cycle to the next with clear and repeatable beat-phase dependent intensity variations. To explain these, we propose the operation of a magnetic valve at L1. The magnetic valve modulates the mass-transfer rate, as evidenced by a factor of 5 variation in orbital-averaged intensity, over the course of the beat cycle in a repeatable manner. The accretion stream threading distance from the WD is also modulated at the beat-period, because of the variation of the WD magnetic field with respect to the stream and because of changes in the mass transfer rate due to the operation of the magnetic valve. Changes in the threading distance result in significant shifts in the position of accreting spots around the beat cycle. As a consequence, only the faintest photometric minima allow for an accurate ephemeris determination. Three regions on the white dwarf appear to receive most of the accretion flow, suggestive of a complex WD magnetic field. 
	
\end{abstract}


\section{Introduction} \label{sec:intro} 

Cataclysmic variables (CVs) are binaries containing a white dwarf (WD) accreting matter from a Roche-lobe filling, M-type, main sequence donor. CVs typically have orbital periods of less than 12-hr and are classified most generally into non-magnetic CVs and magnetic CVs (mCVs). The mCVs are further divided into the lower-field (B $<$ 10 MG) intermediate polars (IPs), which usually contain an accretion disk, and the higher field (B $>$ 10 MG) polars. Polars do not possess an accretion disk due to the strong magnetic field (10-250 MG) of the WD, see the book by \citet{Warner95}, for a review of all types of CVs. In polars, mass transfer takes place from the inter-Lagrangian (L1) point, from which the accretion stream travels ballistically until the magnetic field takes control of the flow of plasma at the threading location. After being threaded, the flow is directed onto the WD at the foot-points of the magnetic field.  

\begin{figure*}
  \includegraphics[width= \linewidth]{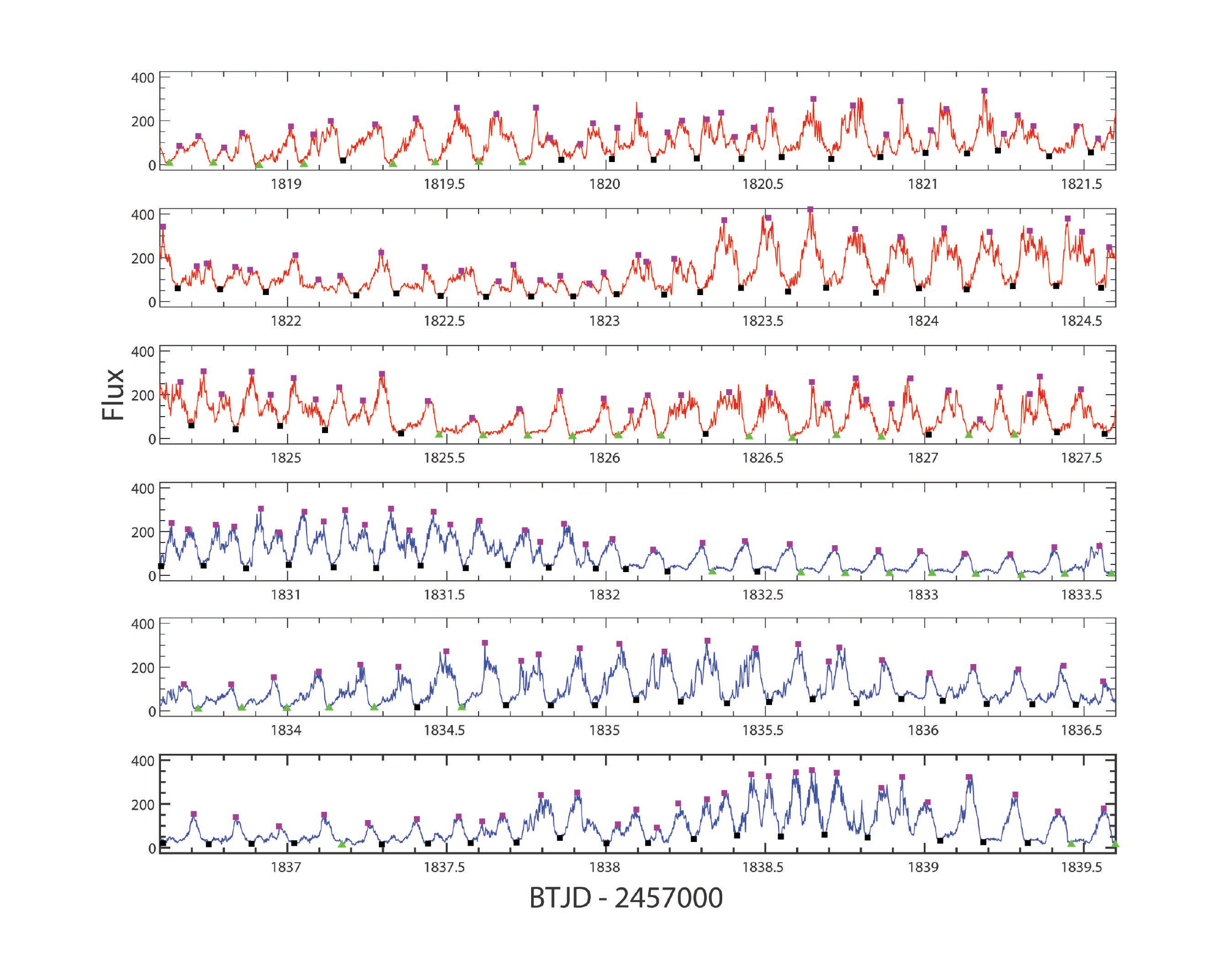}
  \figcaption{Two 9-day data segments are shown to illustrate the nature of the light curve transitions and the measurements performed. Data from the first part of the TESS observation are shown in red and data from the second part are shown in blue. One or two peaks occur during each orbital cycle and have been labeled with purple squares. Most of the light curve minima are shown as black squares, while the faintest minima are shown as green triangles, as they are the most reliable markers for ephemeris determination. Notice that every few days there is a significant change in the light curve due to spin-orbit asynchronism.
 }
\label{fig:3Days}
\end{figure*}
      
BY Camelopardalis (hereafter BY Cam) is the prototype of a small sub-class of polars known as asynchronous polars (AP). In the vast majority (96$\%$) of polars, the magnetic field is sufficiently strong such that magnetic locking of the binary occurs. Hence, the WD rotates at the same rate as the binary orbital period. In APs this is not the case. Instead, the WD rotates a few percent faster, or slower, than the binary. There are only eight confirmed APs corresponding to $\sim$6$\%$ of the known polars, see Table \ref{table:APs} 
for a summary of their properties. Recent TESS observations of CD Ind \citep{Hakala19, Littlefield19, Mason20} show it to be a short-period analog of BY Cam. 

\begin{figure}
\includegraphics[width= \linewidth ]{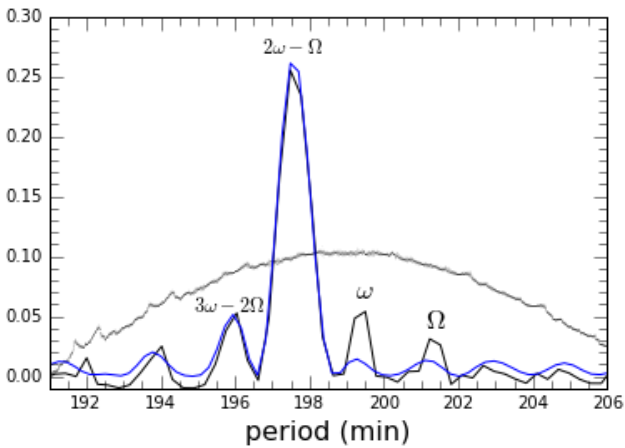}
 \figcaption{The Phase Dispersion Minimization (PDM: black) and Lomb-Scargle (LS: blue) power spectrum of the TESS light curve. The PDM statistic is inverted to match the LS power. The PDM method is applied in two ways. The broad parabolic curve is the 1-day averaged PDM which identifies the spin period, but lacks precision. The labeled black line is the PDM of the full Sector and matches the LS power in blue, except that the PDM also able to detect the spin frequency, $\omega$,  while the LS method detects only a weak signal. We adopt the TESS PDM results for the side-band and spin periods and the spectroscopic orbital period, $\Omega$, of \citet{Schwarz05}. The labeled periods are 195.927, 197.560, 199.384, and 201.244 minutes respectively. Most of the signal is at the side-band frequency, 2$\omega$ - $\Omega$.  
\label{fig:PDM_LS}}
\end{figure}

\begin{deluxetable*}{ccccccccc}
\tablecaption{The asynchronous polars. \label{table:APs}}
            
\tablehead{
\colhead{Name} &
\colhead{$P_{spin}$ (m)} &
\colhead{$P_{orb}$ (m)} &
\colhead{$P_{spin}/P_{orb}$} &
\colhead{$P_{beat} (d)$} &
\colhead{distance (pc)} &
\colhead{M1 (M$\odot$)} &
\colhead{B (MG)} &
\colhead{references}
}
\startdata
\igr      & 81.07 & 83.4 & 0.972 & 2.0 & $165.5^{+1.9}_{-1.5}$ & & $<$20 & (1) \\
\rxs      & 94.46 & 98.4 & 0.96 & 1.8 & $156.0^{+1.9}_{-2.3}$ & & & (2)  \\
CD Ind    & 110.97 & 112.2 & 0.989 & 7.0 & $242.2^{+5.9}_{-5.6} $ & 0.87 & 11 & (3,4,5) \\
Paloma    & 136.76 & 157.2 & 0.87 & 0.73 & $582^{+28}_{-20}$ & & & (6) \\
V1500 Cyg & 198.24 & 201.06 & 0.986 & 9.83 & $1567^{+270}_{-192} $ &   & & (7) \\
BY Cam    & 199.384 & 201.244 & 0.991  & 14.26  & $264.5^{+1.9}_{-1.6}$ & 0.76 & 41, 168 & (this work, 8, 9, 10, 11) \\
V1432 Aql & 202.35 & 201.96 & 1.002 & 70 & $449.7^{+6.8}_{-6.5}$ & & & (12)\\
\fullname & 270.6 & 278.4 & 0.972 & 6.7 & $1233^{+796}_{-286}$ & & & (6)\\
 \enddata
            
\tablecomments{All distances are the geometric distances computed by \citet{BJ20} from Gaia EDR3 \citep{edr3} parallaxes. References (1) \citet{Tovmassian17}, (2) \citet{halpern}, (3) \citet{Littlefield19}, (4) \citet{Dutta2022}, (5) \citet{Schwope97},
 (6) \citet{Littlefield22}, (7) \citet{Pavlenko18}, (8) \citet{Shaw20}, (9) \citet{Cropper89}, (10) \citet{Tutar17}, (11) \citet{Schwarz05}, (12) \citet{Littlefield15}.}
\end{deluxetable*}

 \begin{figure*}
  \includegraphics[width=\textwidth]{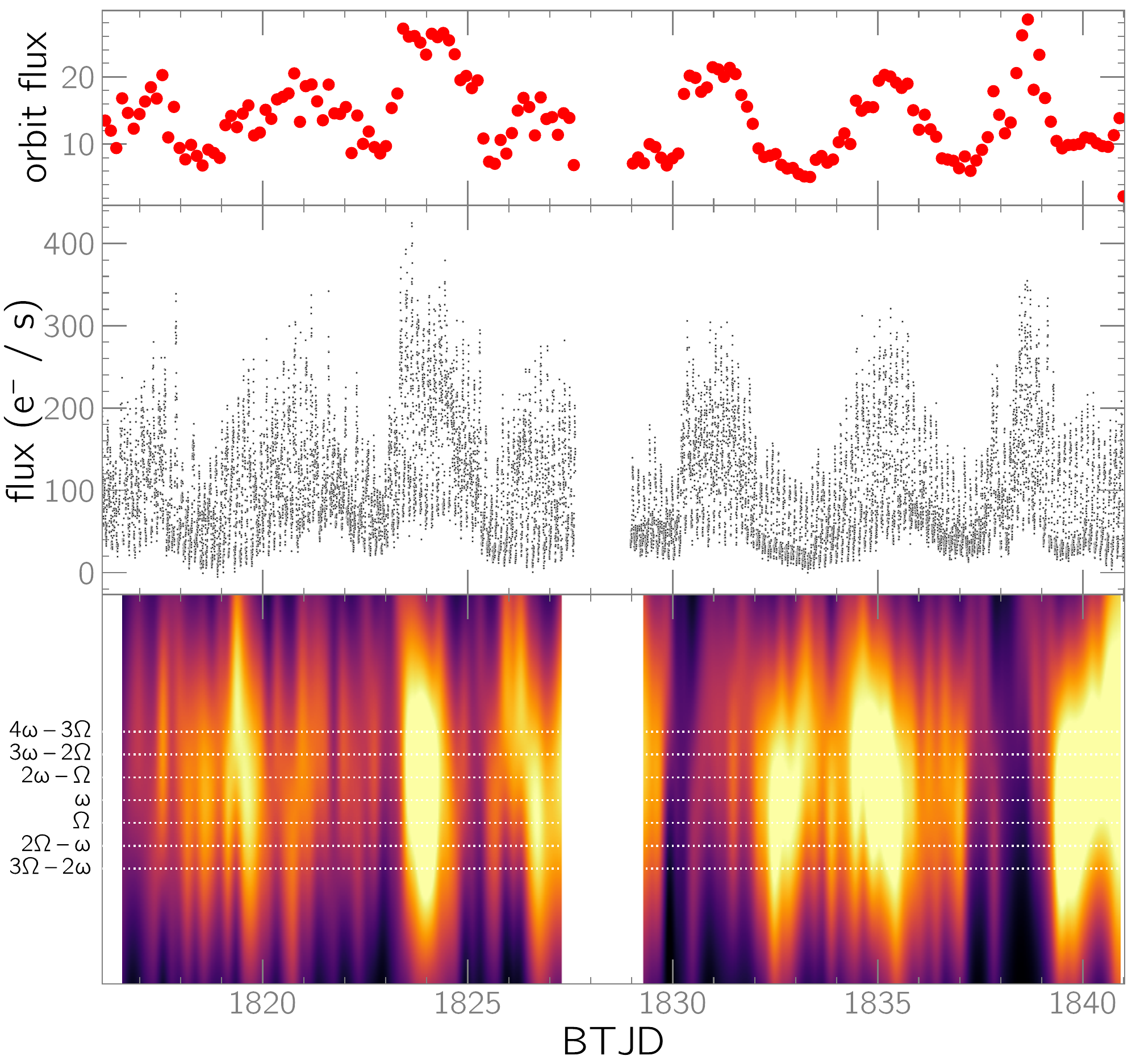}
  \figcaption{The TESS light curve of BY Cam. The data are in two parts of about 11 days each with a 3 day gap in between. ({\bf Top}) The orbit-averaged brightness shows a factor of 5 - 10 variation over the beat cycle. The brightest peaks are separated by the beat period of 14.26 days.  ({\bf Middle}) The light curve, obtained at 2-min cadence, is shown. The individual spikes represent variations due to the spin of the WD. ({\bf Bottom}) The trailed (2D) Lomb-Scargle power spectrum is shown using a trailing window of 1 day. The WD spin frequency, $\omega$, is only detected when the light curve exhibits stability between pole-switching, especially near dates 1824, 1835, and 1840. Some accretion spot drifting is seen as a shift in frequency, punctuated by intervals of stability, before 1835. At times of magnetic-pole switching, e.g. near dates 1825, 1830, and 1838, no periodic signal is observed.
  \label{fig:2D}}
  \end{figure*}

BY Cam (H0538+608) was discovered using the {\it{High-Energy Astrophysics Observatory-1}} (HEAO-1) \citep{Remillard86}. Quickly, peculiarities became apparent. It exhibits both positive and negative circular polarization ($\sim$10$\%$) and photometric light curves that are similar to synchronous polars, except that very little linear polarization is observed \citep{Mason87, Mason89}. The source displays emission-line characteristics that are similar to synchronous polars, like AM Her. Strong Balmer lines and a high HeII/H$\beta$ ratio \citep{Remillard86, Mason89} are observed and are indicative of an mCV. However, BY Cam also shows unusually rich emission line profiles displaying not only the usual broad and narrow components, due to the main accretion funnel just above the white dwarf's surface and the irradiated face of the companion respectively, but also shows two additional emission-line components, including an extremely high-velocity component \citep{Mason96, Mason89}. The emission lines of BY Cam are also peculiar in that accreting material has an unusually high N/C line ratio \citep{Bonnet-Bidaud87, Mouchet03} and show complex emission-line periodicity variations \citep{Zucker95, Mouchet97}. Optical Doppler tomograms suggest that the Balmer emission originates near the donor star, while the HeII emission is from close to the WD \citep{Szkody00}.

BY Cam is a member of group of polars known for their unusually hard X-ray emission and XMM-Newton observations of BY Cam show both hard and soft X-ray poles \citep{Ramsay02}.  Generally speaking, at any one time BY Cam displays X-ray, UV, optical, IR, and radio properties indistinguishable from a synchronous polar. However, when observations are made for a time greater than a few orbital cycles, the larger baseline observations display accretion geometry changes due to spin-orbit asynchronism. However, its high hard X-ray flux compared to other polars, as measured by the INTEGRAL/IBIS telescope, suggests that the hard X-ray emission properties of BY Cam and APs in general have similarities to those of the IPs. \citep{Scaringi10}. Motivated by the discovery of asynchronism of the AP V1500 Cyg, apparently the result of the eruption Nova Cyg 1975, \citet{Pagnotta16} searched for nova shells surrounding BY Cam, CD Ind, and V1432 Aql and found none. 

Several long-term studies of polarization and photometry have demonstrated that changes in the accretion flow onto the white dwarf in BY Cam occurs on time-scales of a few days; however, accretion flow structure repeats on time-scales of weeks to months and even years. Spectroscopic observations of the narrow emission-like component originating on the heated face of the donor, revealed an orbital period that a few percent longer than the main photometric period \citep{Silber92}. The spin period was derived from polarization as well as optical and X-ray photometry \citep{Mason89, Mason95, Mason96,  Ramsay96, Silber97, Mason98}. Thus, it became clear that BY Cam must be an AP. However, several photometric/polarization and spectroscopic periods are detected and the correct identification of the white dwarf spin and orbital periods remained controversial \citep{Piirola94, Honeycutt05, Wang20}.  Extensive spectroscopy was used to derive a precise orbital period \citep{Schwarz05}, yielding P$_{orb}$ = 201.244 min. Even more comprehensive photometric campaigns were able to establish the time evolution of both the dominant photometric signal \citep{Andronov08, Pavlenko13, Babina19} and the proposed spin period of the white dwarf.

A confounding factor in establishing the correct identification for the spin period of the WD is the complex nature of the light-curve variability, beyond its asynchronism, which is attributed to a complex magnetic field of the white dwarf \citep{Mason95, Wu96, Mason98, Zhilkin12, Pavlenko13, Zhilkin16}. While a simple dipole field structure was found to be incompatible with polarization and photometric observations, many details remain uncertain. The high-cadence continuous photometry, allowed by TESS, covers the beat cycle of BY Cam for the first time and places the strongest constraints on models thus far.

In polars, including BY Cam, the optical light is dominated by cyclotron emission from the accretion column(s) or funnel(s) located just above foot-points of magnetic field lines. V1500 Cygni is an exception, where the heated face of the donor still dominates the optical emission \citep{Pavlenko18}, due to irradiation by the hot WD, as it cools after the 1975 nova eruption. In APs, the near-WD field connects the magnetic foot-points with the stream threading region \citep{Mukai88} located in the orbital plane. 
 
\section{Observations} \label{sec:style}
BY Cam was monitored by TESS during Sector 19, between 27 November 2019 and 24 December 2019 at 2-min cadence. The TESS band-pass is wide ($\sim$600-1,000~nm), extending across both the $I$ and $R$ bands and into the near-IR, see \citet{Ricker2015} for a description of TESS. Data was downloaded using the Python package {\tt lightkurve} \citep{lightkurve}. Since BY Cam is comparatively bright, and uncrowded, there were no issues with photometric reduction.  The data consists of two parts, separated by a short gap, as shown in separate panels of Figure \ref{fig:3Days}. Light-curve peaks and minima are labeled in Figure \ref{fig:3Days} for analysis. Part one of the data, the red curve, consists of about 11 days and thus covers about 75$\%$ of the $\sim$ 14-day spin-orbit beat cycle. There is an interruption for about two days due to poor data during the perigee crossing of TESS. The data collection is continued for another 11 days, that we call part two, shown as the blue curve in Figure \ref{fig:3Days}. As a result of the gap, about 8$\%$ of the beat cycle is not covered by TESS, but about 71$\%$ of the beat cycle is covered twice.

Extreme brightness variations are seen over the course of just a few days. One can easily see frequent transitions in the light curve structure occurring in Figure \ref{fig:3Days}, indicating changes in the accretion geometry - the flow structure and/or impact position on the WD.
We focus now on part 2 shown in the bottom panel of Figure \ref{fig:3Days}, for which the light curve is somewhat more stable than in part 1. Around day 1832 (blue curve: Figure \ref{fig:3Days}), the mean intensity decreases by a factor of $\sim$ 5 and the geometry transforms from a two-pole state to a one-pole accretion configuration. At about date 1834, more complex changes occur over a period of many cycles, reaching a bright blended state. By date 1836, single-poles are again observed, followed by a complex transition to a brighter two-pole state between dates 1838 and 1839, after which a single-pole configuration is again observed until the end of part two.

\begin{figure*}
\centering
 \includegraphics[width = 6.1in]{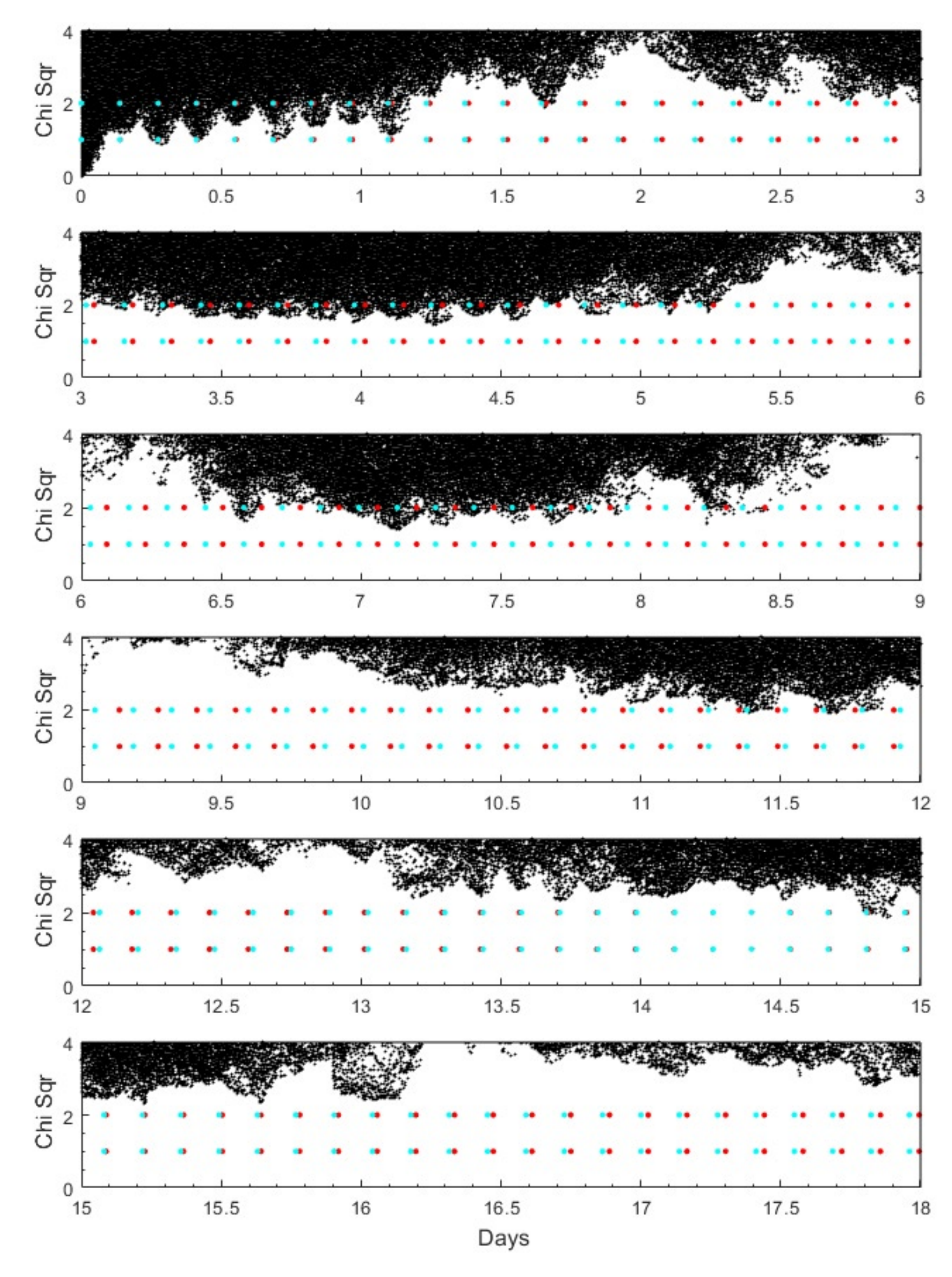}
 \vspace{-.15 in}
  \figcaption{The cross-correlation coefficient was calculated for randomly chosen pairs of data snippets, then the reduced chi-squared is shown as a function of the snippet time separation. Multiples of the spin period (red dots) and side-band period (blue dots) are shown for comparison. Two sets of these dots are placed to guide the eye in noticing both strong and weak correlations. The top panel shows how light curve snippets separated by 1 day or less measure the WD spin period, while longer time separations do not. The correlation strengthens at 7.13-days, roughly one-half of the beat cycle. Weaker correlations in the fifth panel suggest a beat-period of 14 - 15 days. Significant variations between data obtained at successive beat phases make a more precise beat-period measurement impossible.
  \label{fig:CC}}
\end{figure*}

\begin{deluxetable}{cccccccc}'
\tabletypesize{\scriptsize}
\tablecaption{Tess Light Curve Minima}
\label{tab:table_2}
\tablenum{2}
\tablehead{\colhead{BTJD} & \colhead{Flux} & \colhead{BTJD} & \colhead{Flux} & \colhead{BTJD} & \colhead{Flux} & \colhead{BTJD} & \colhead{Flux} } 

\startdata
16.1800 & 32 & 22.2160 & 28 & 29.4820 & 26 & 35.2340 & 42 \\
16.3254 & 29 & 22.3420 & 37 & 29.6217 & 19 & 35.3795 & 34 \\
16.4562 & 19 & 22.4810 & 26 & 29.7590 & 17 & 35.5125 & 40 \\
16.6003 & 35 & 22.6235 & 21 & 29.8953 & 16 & 35.6490 & 53 \\
16.7390 & 21 & 22.7650 & 21 & 30.0341 & 16 & 35.7865 & 35 \\
16.8724 & 18 & 22.8969 & 23 & 30.1715 & 21 & 35.9270 & 54 \\
17.0030 & 26 & 23.0330 & 32 & 30.3150 & 48 & 36.0570 & 45 \\
17.1485 & 33 & 23.1820 & 31 & 30.4630 & 34 & 36.1951 & 31 \\
17.2941 & 37 & 23.2955 & 42 & 30.6030 & 41 & 36.3381 & 30 \\
17.4288 & 38 & 23.4235 & 62 & 30.7375 & 45 & 36.4741 & 28 \\
17.6554 & 21 & 23.5717 & 44 & 30.8698 & 32 & 36.6097 & 21 \\
17.7943 & 27 & 23.6903 & 63 & 31.0047 & 48 & 36.7529 & 16 \\
17.9450 & 24 & 23.8475 & 40 & 31.1451 & 36 & 36.8870 & 17 \\
18.0907 & 13 & 23.9820 & 59 & 31.2790 & 34 & 37.0219 & 21 \\
18.2100 &  0 & 24.1325 & 55 & 31.4185 & 45 & 37.1713 & 13 \\
18.3520 & 13 & 24.2770 & 70 & 31.5593 & 34 & 37.2970 & 16 \\
18.4980 &  6 & 24.4125 & 70 & 31.6925 & 48 & 37.4410 & 18 \\
18.6295 &  6 & 24.5545 & 63 & 31.8198 & 36 & 37.5751 & 22 \\
18.7675 &  6 & 24.6985 & 59 & 31.9685 & 31 & 37.7197 & 25 \\
18.9117 &  5 & 24.8370 & 41 & 32.0621 & 28 & 37.8553 & 44 \\
19.0520 &  0 & 24.9769 & 56 & 32.1930 & 17 & 38.0010 & 19 \\
19.1750 & 19 & 25.1180 & 38 & 32.3341 & 15 & 38.1317 & 22 \\
19.3314 &  1 & 25.3573 & 23 & 32.4747 & 16 & 38.2751 & 40 \\
19.4639 &  8 & 25.4760 & 15 & 32.6121 & 11 & 38.4112 & 55 \\
19.6010 & 11 & 25.6143 & 12 & 32.7510 &  8 & 38.5490 & 51 \\
19.7379 &  8 & 25.7541 & 11 & 32.8900 &  8 & 38.6860 & 60 \\
19.8595 & 21 & 25.8949 &  8 & 33.0237 &  9 & 38.8210 & 47 \\
20.0185 & 25 & 26.0390 & 12 & 33.1610 &  4 & 39.0490 & 32 \\
20.1499 & 21 & 26.1731 & 12 & 33.3041 &  0 & 39.1849 & 26 \\
20.2845 & 29 & 26.3132 & 20 & 33.4401 &  5 & 39.3240 & 22 \\
20.4255 & 26 & 26.4495 &  7 & 33.5865 &  6 & 39.4605 & 15 \\
20.5510 & 33 & 26.5851 &  0 & 33.7201 &  7 & 39.5985 & 14 \\
20.7063 & 26 & 26.7230 & 13 & 33.8570 & 13 & 39.7370 & 22 \\
20.8610 & 33 & 26.8643 &  5 & 33.9980 & 12 & 39.8750 & 22 \\
21.0030 & 53 & 27.0130 & 17 & 34.1310 & 13 & 40.0129 & 24 \\
21.1335 & 51 & 27.1390 & 13 & 34.2725 & 14 & 40.1527 & 25 \\
21.2300 & 64 & 27.2825 & 15 & 34.4070 & 16 & 40.2905 & 18 \\
21.3910 & 38 & 27.4153 & 28 & 34.5470 & 15 & 40.4261 & 16 \\
21.5220 & 56 & 27.5650 & 20 & 34.6862 & 26 & 40.5710 &  9 \\
21.6553 & 59 & 29.0697 & 27 & 34.8267 & 26 & 40.7060 &  4 \\
21.7890 & 55 & 29.2070 & 20 & 34.9655 & 27 & 40.8471 & 18 \\
21.9319 & 43 & 29.3488 & 20 & 35.0941 & 50 & 40.9843 &  7 \\
\enddata
\end{deluxetable}

\subsection{Time-Series Analysis} \label{subsec:LC}

In order to investigate photometric periodicity, we performed time-series analysis of the TESS data using several independent methods. We used 1D and 2D Lomb-Scargle (LS) power spectra \citep{Lomb76, Scargle82}, Phase Dispersion Minimization (PDM) \citep{Stellingwerf78}, and a Cross-Correlation (CC) analysis. First LS and PDM searches were performed on the entire TESS dataset and the results are shown in Figure \ref{fig:PDM_LS}, where the PDM periodogram is inverted vertically to match the LS power. The WD spin period, 199.384-min, is labeled as the frequency $\omega$ and the orbital period, 201.244-min, is labeled with the frequency $\Omega$. Both the PDM and the LS power spectrum peak at the side-band frequency, labeled 2$\omega$ - $\Omega$ corresponding to the side-band period of 197.560 min seen many times in BY Cam photometry \citep{Silber97, Mason98, Honeycutt05}.

An alternative WD spin identification model exists, where the 197.560-min period is ascribed to the WD spin. This is a natural proposal since the dominant signal in the power spectrum of polars is most often the WD spin period. This is true for BY Cam also, but only when the period analysis is performed while the accretion geometry remains stable, which as we shall demonstrate in the next subsection, is about 1-day. We adopt the period identifications given in Figure \ref{fig:PDM_LS} and  Table \ref{table:APs} for BY Cam based on X-ray and polarization studies \citep{Mason89, Mason98} for the WD spin, and emission-line spectroscopy for the orbital period \citep{Mason96,Schwarz05}.

Notice the differences in the PDM versus the LS results. The PDM picks up stronger signals at both the proposed spin, $\omega$, and orbital $\Omega$, periods. This is due to destructive interference in the LS frequency analysis, from to pole switching to opposite sides of the WD. The alternative WD spin period assignment, with $\omega$ being the strongest peak, is challenged to explain the appearance of the other peaks in the PDM of Figure \ref{fig:PDM_LS}.

A third interpretation has been presented \citep{Honeycutt05, Wang20} in which the 197.560-min period (the dominant signal: Figure \ref{fig:PDM_LS}) is the binary orbital period. The argument for this assignment is the apparent long-term stability in that period observed by \citet{Honeycutt05}. However, a comparison of decades of independent period determinations show a clear progression towards longer photometric periods over time, e.g. \citep{Mason89, Silber97, Mason98, Honeycutt05, Andronov08, Pavlenko13, Babina19, Wang20} and this work.

In order to capture the complex time evolution inherent in the BY Cam light curve a 2D-LS period analysis is performed, which may also be described as a temporally resolved or trailed power spectrum and is shown as a contour plot in Figure \ref{fig:2D}. The top panel shows the orbit averaged light curve, showing up to a factor of 6 total flux variation, while the middle panel shows the full light curve. In the trailed power spectrum (bottom) a sliding window of 1 day was selected to match the time-scale for light-curve stability. The many vertical stripes are indicative of sporadic period stability. Light-curve morphology is most stable near days 1824 and 1840. It becomes aperiodic between 1821 and 1823, near 1825, near 1830, and again between 1837 and 1839. Aperiodic intervals indicate changes in the flow onto the WD, thus modifying accretion geometry. These power-spectrum features repeat on the spin-orbit beat cycle.
Periodic high-intensity episodes correspond to the previously identified X-ray flaring state \citep{Ishida91}, in contrast to the single-pole, pulsed state. Two-pole accretion (where two spots located on the same hemisphere are active) and the system is much brighter, are seen at various times in Figure \ref{fig:3Days} and in the top panel of Figure \ref{fig:2D}.  The higher amplitude variations are correlated with higher orbital averaged flux values (Figure \ref{fig:2D}: compare the top and middle panels).

\subsection{Cross-Correlation Analysis}
\label{subsec:CC}

In order to search for periodicity in an independent manner that also sheds light on the coherence time of the BY Cam light curve, a cross-correlation analysis is performed. It involves clipping a random section of 4.5-hrs of continuous data that we call a snippet. The cross-correlation coefficient was calculated for randomly chosen pairs of snippets. The cross-correlation is a measure of snippet  similarity. The reduced $\chi$-squared of the cross-correlation results are shown as a function of snippet time separation in  Figure \ref{fig:CC}, 
using one million random pairs of data snippets. Whenever the $\chi$-squared comes down near 1 in Figure \ref{fig:CC}, we find repeating light-curve shapes. Looking at the left-hand side of the top panel of  Figure \ref{fig:CC} we see several significant dips at equal intervals until about 1.1-days. This corresponds to the timescale for accretion structure stability, see also \citet{Mason20}. If the snippets are separated by more than 1.1-day, then the snippet light curves do not correlate well. Snippets separated by less than 1-day have a high likelihood of a strong correlation when phased at some multiple of the WD spin period. 

In order to investigate the spin and side-band periods, two phase clocks are represented as blue and red dots in  Figure \ref{fig:CC}. 
The red dots start at 0 and occur at multiples of the spin period, P$_{spin}$ = 199.384-min. The blue dots also start at 0 and track the shorter side-band period, P$_{sb}$ = 197.560-min.  Unfortunately, by the time the red and blue dots diverge, the light curve has lost its coherence. Looking again at the top panel of  Figure \ref{fig:CC}, 
the light curve completely changes shape after 1.1-days. This is easily verified by examination of the light curves in  Figure \ref{fig:3Days}.

This WD spin model degeneracy is well known and discussed in some detail by \cite{Mason20} who describe the case for the same effect in CD Ind, a short period AP. Also see \cite{Hakala19} and \cite{Littlefield19} for TESS studies of CD Ind. Looking at  Figure \ref{fig:CC}, it is not possible to refute this alternative model for BY Cam on the basis of the CC analysis. However, the reason for the dominant photometric signal being associated with the side-band signal is clear. Starting about 3.5-days (weak) periodic signals appear, see the second panel from the top of  Figure \ref{fig:CC}. The strongest CC signal, after 1.1-days, is at 7.13-days (third panel) and is identified as one-half of the spin-orbit beat cycle in the preferred model, giving a beat period of 14.26-days. At this point, according to the preferred model, the accretion flow has been transferred to the other side of the WD. In the alternative model both one- and two-pole accretion occurs onto only one hemisphere of the WD throughout the entire beat cycle, 7.13-days in this case, and is not excluded by the TESS data alone, however see the analysis of BY Cam in \citet{Mason20}. With that said, we stress that the results of the current study do not depend on the correct identification of the WD spin period in BY Cam.

\subsection{White Dwarf Spin and Side-band Ephemerides}

The PDM and LS results (Figure \ref{fig:PDM_LS}) are used to determine the WD spin period for the TESS ephemeris. The zero-point of the ephemeris is derived from the lowest flux minima, namely those with less than 15 flux units and those are displayed as green triangles in Figure \ref{fig:3Days}. The TESS WD spin ephemeris for minima, is thus: 
$$    T_{BTJD} = 2458827.2742(20) + 0.1384611(70)E, $$
where $E$ is the integer cycle count. This is not a long-term ephemeris for two reasons: 1) the TESS observation is about 27-days, so is capable of a period determination of only about 1-s uncertainty and 2) the WD spin period of BY Cam is changing, so a high-precision long-term linear ephemeris is not appropriate. Note that this period is a full 3-s longer than the original high precision ephemeris \citep{Mason89}. The intrinsic variability of BY Cam makes period determinations difficult. However our finding that lower intensity minima are the most reliable, to be addressed in the discussion section, will aid future non-linear ephemeris determinations. Measurements of light-curve minima, shown in  Figure \ref{fig:3Days}, are listed in Table \ref{tab:table_2}. 
The TESS side-band ephemeris, for minima, is: 
$$    T_{BTJD} = 2458833.1654(1) + 0.1371944(70)E. $$
Either of these ephemerides may be used to predict light-curve minima, especially those at lower flux levels. 
However, the side-band ephemeris produces more scatter.

\begin{figure}
\centering
\includegraphics[width= 1.65 in]{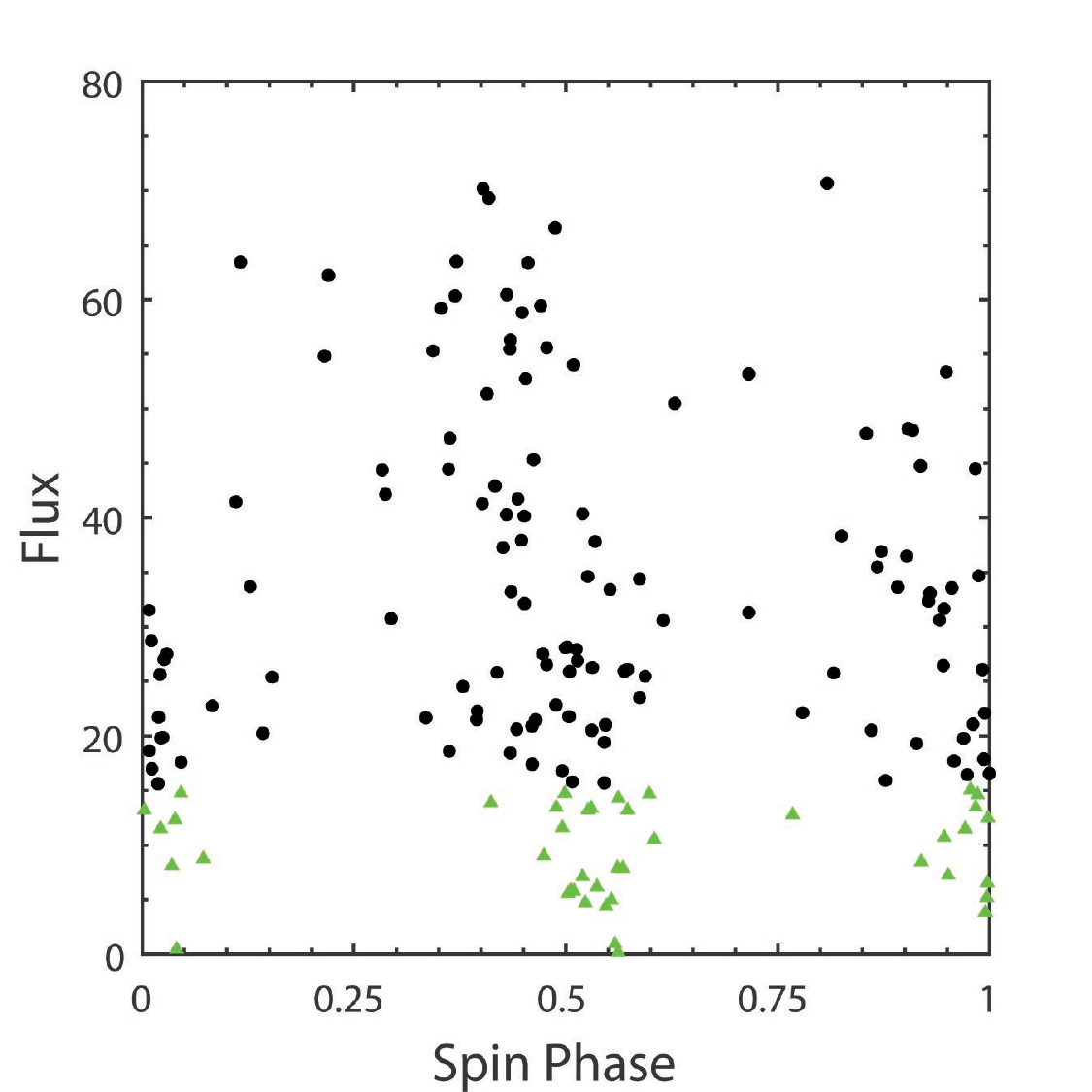}
\includegraphics[width= 1.65 in]{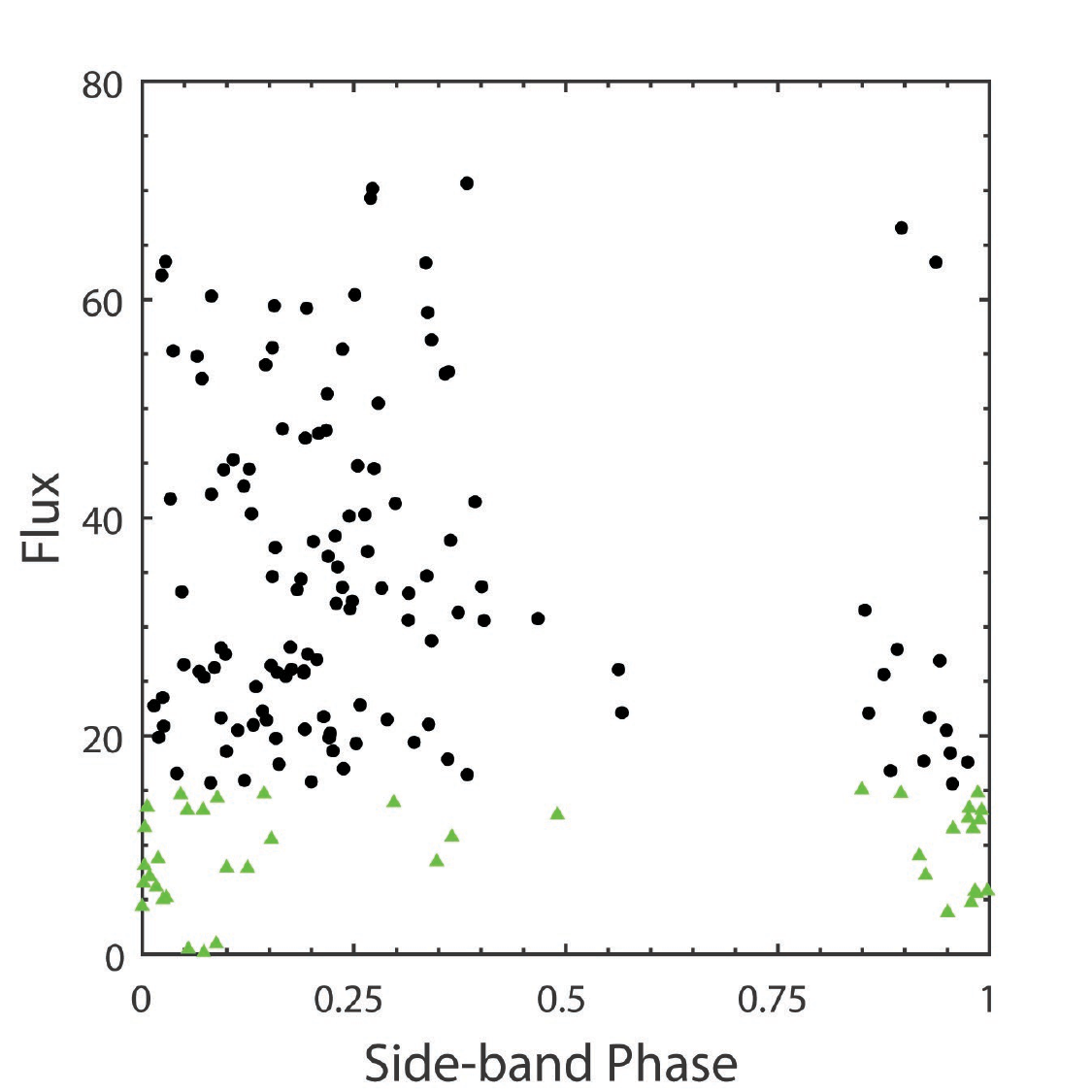}
  \figcaption{TESS light curve minima, shown in  Figure \ref{fig:3Days} are phased with the spin (left) and side-band (right) photometric ephemeris respectively. The minima exhibit less dispersion at low flux levels, shown as green triangles here and in Figure \ref{fig:3Days} {\bf {Left:}} The minima are phased with the white dwarf spin period of 199.384-min. The points clearly divide into two groups, one at spin-phase 0.0 and the other near phase 0.5. The first very tight clump of green triangles, at phase 0.05 in this panel, is used for the zero point of both ephemerides. {\bf {Right:}} The same minima now are phased with the side-band period of 197.560-m. The dispersion at low flux (green triangles) is larger than those phased with the proposed spin period. At high flux levels (black squares) the dispersion increases.
\label{fig:PhasedPeriods}}
\end{figure}

\begin{figure}
  \includegraphics[width=1.65 in]{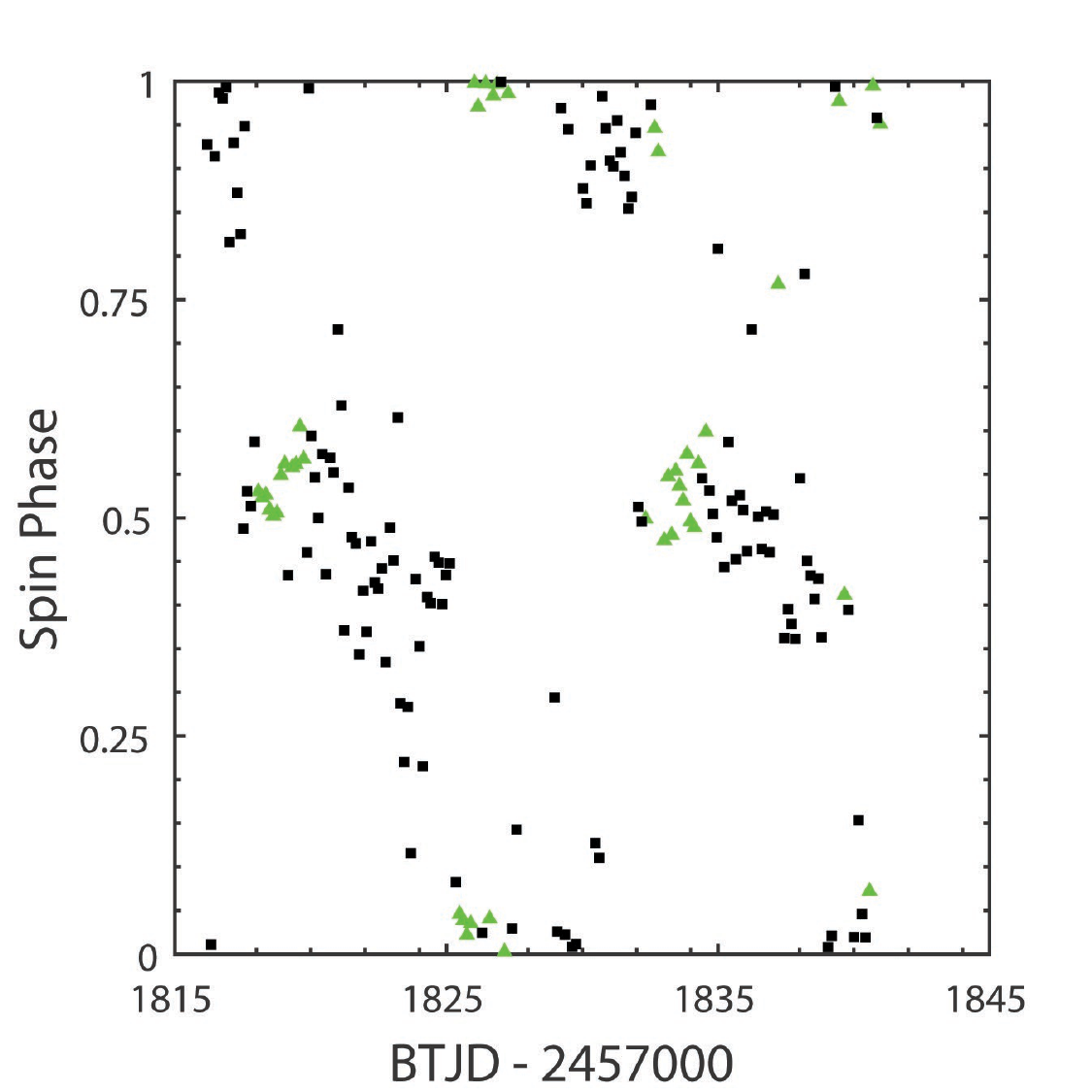}
 \includegraphics[width=1.65 in]{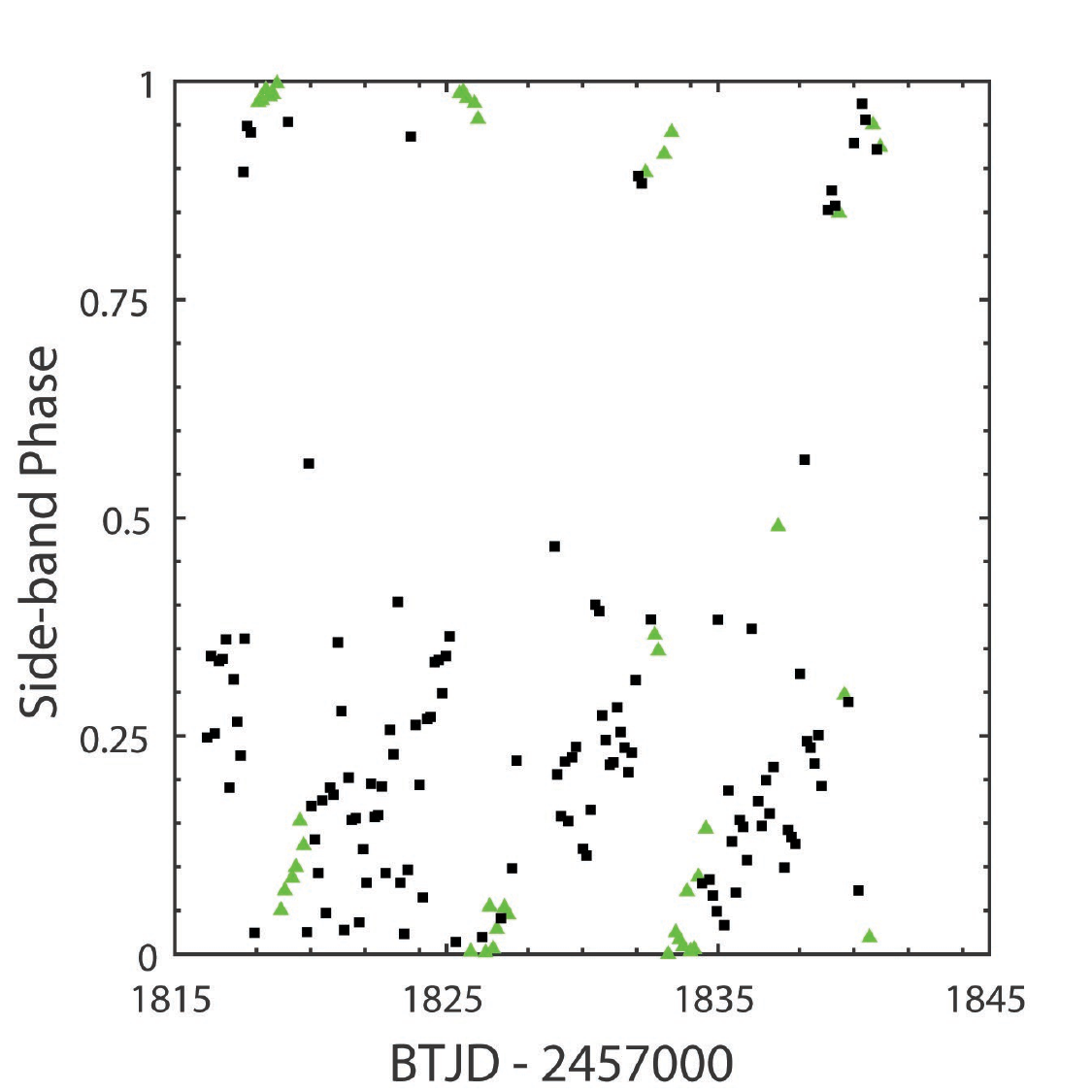}
  \figcaption{The minima shown in  Figure \ref{fig:3Days} are now shown with spin and side-band phase versus time. {\bf {Left:}} Accretion pole switching is clear at the lower flux levels (green triangles). At higher fluxes (black squares) some drifting in phase over time occurs. {\bf {Right:}} Light curve minima are now phased using the side-band period. Substantial drifting is seen, suggesting that the side-band period is not likely tracking the WD spin.
\label{fig:Phase_Vs_Time_min}}
\end{figure} 

\begin{figure}
  \includegraphics[width=1.65 in]{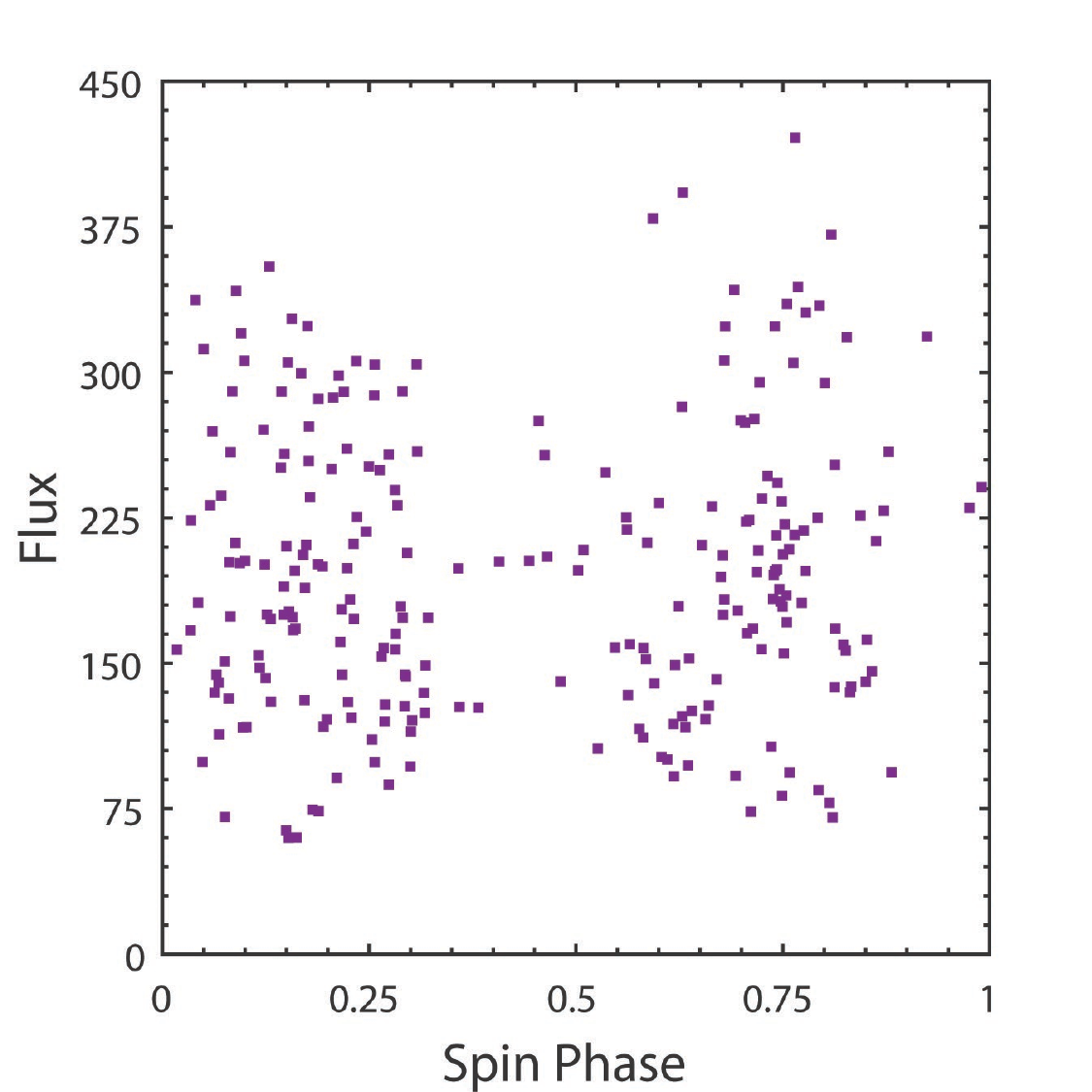}
 \includegraphics[width=1.65 in]{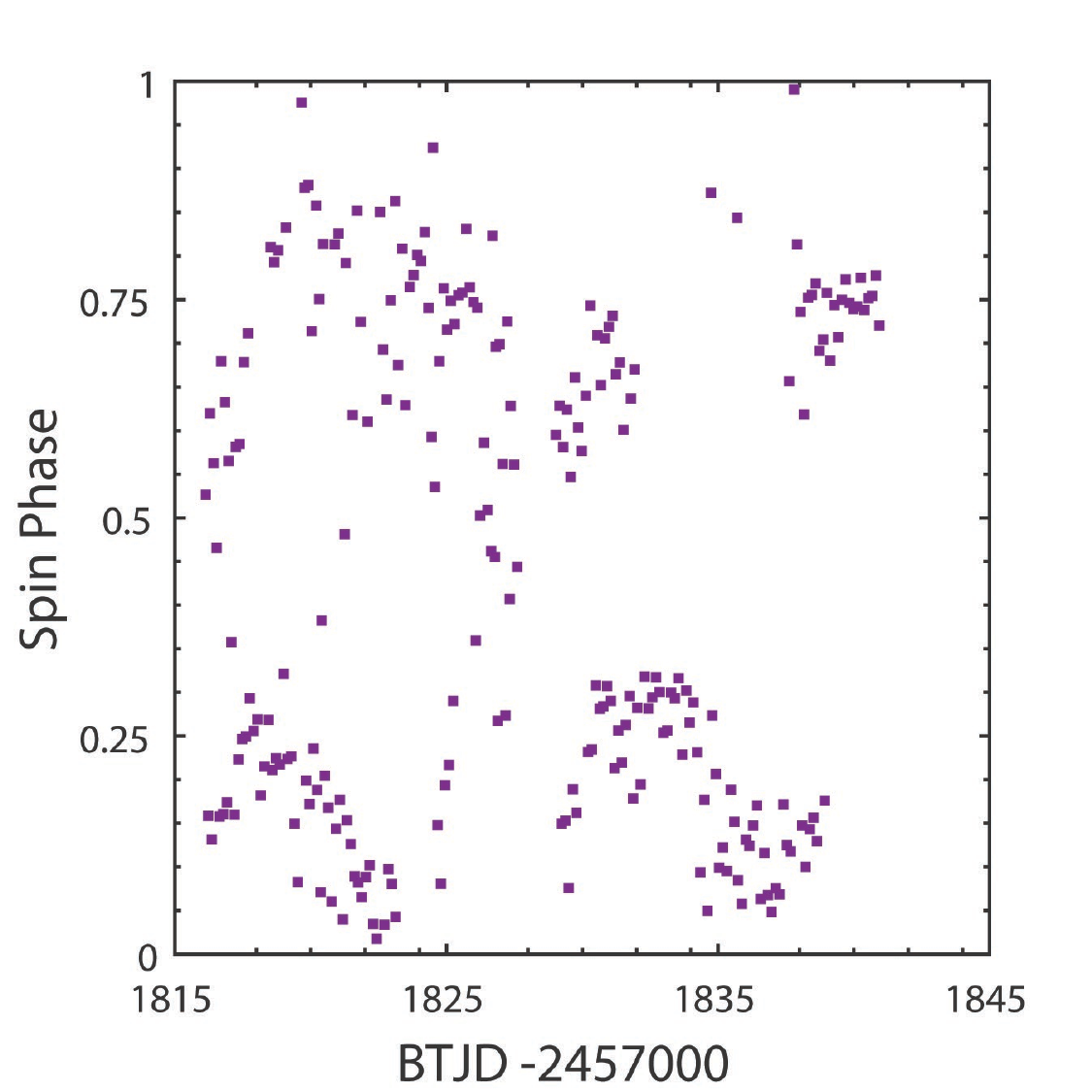}
  \figcaption{Light curve maxima, from  Figure \ref{fig:3Days} are phased with the WD spin period. One maximum is measured when there is a one-pole configuration, while two maxima are measured for two-pole configurations. {\bf Left:} Relative flux of maxima are plotted vs. spin phase and show two clumps of points with large scatter in both intensity and phase. A single larger clump is observed when maxima are phased with the side-band period (not shown).  {\bf Right:} When the spin phase is shown as a function of time, pole switching is seen to occur as well as phase drifting, e.g. near days 1820 and 1835. These transitions are also seen in the 2D power spectrum of Figure \ref{fig:2D}. We propose that this phase drifting is due to a changing threading distance, resulting from the operation of the magnetic valve.
  \label{fig:Phase_Vs_Time_max}}
\end{figure}

Photometric minima have been used for timing analysis of BY Cam because they maintain relative stability and a distinctive minimum is observed during practically every cycle, while photometric maxima are highly variable. However, the minima have a large range in observed intensity. In order to examine the phase dependence of the measured minima, recall Figure \ref{fig:3Days}, we phased them first with the spin period, shown in the left panel of  Figure \ref{fig:PhasedPeriods} and then with the side-band period shown in the right panel of Figure \ref{fig:PhasedPeriods}. With the minima phased with the WD spin period, left panel, two clusters of points are observed. Importantly, the scatter within the clusters is significantly reduced at low flux levels. The same is true for the side-band phased minima (right panel). However, in the right panel the scatter in phase at all flux levels is higher than in the left panel. 

To examine spin phase and side-band phase evolution during the full TESS observation, minima are plotted in Figure \ref{fig:Phase_Vs_Time_min}. Pole shifting is apparent in the left panel especially at low flux levels, notably the four tight clumps of green triangles. Corresponding pairs of clumped green triangles are seen at the same spin phases, 0.05 and 0.55, on consecutive beat cycles  At high fluxes (black squares) substantial drifting occurs, e.g. between 1815 and 1825. On the other hand, the side-band phased minima constantly drift and/or shift in phase over time, at all flux levels, see the right hand panel of Figure \ref{fig:Phase_Vs_Time_min}. 

 Figure \ref{fig:Phase_Vs_Time_max} shows the maxima, first plotted in Figure \ref{fig:3Days}, now shown phased with the spin period of the WD. The peak is highly variable in both flux and phase and is grouped into two clumps situated roughly 0.5 apart in phase. Phase variations are not random however, as Figure \ref{fig:Phase_Vs_Time_max} (right panel) shows phase drift and accretion pole switching.

\subsection{Orbital Inclination and Spot Positions}
\label{subsec:spots}

The inclination of BY Cam is estimated using the amplitude of the narrow emission-line component, K$_2$ = 210 km/s \citep{Mason89, Mouchet97, Schwarz05}, the WD mass, M$_1$ = 0.76 M$_\odot$ \citep{Shaw20}, and the donor mass, M$_2$ = 0.2 M$_\odot$ \citep{Knigge11}. The narrow emission-line component is highly variable in strength and disappears at orbital phase 0, indicating a relatively large inclination or a small emission region, assumed to be near L1. The radial velocity variations of \citet{Mason89} are non-sinusoidal suggesting that light is shifted from the binary center of mass. Therefore, weighting these factors, we adopt a K-correction \citep{Wade88} of f = +0.35 +/- 0.1, which increases the measured K$_2$ from 210 km/s to 243 +/- 20 km/s. Using the binary mass function to solve for the inclination yields, i = 43$^o$ +/- 8$^o$.

\begin{figure}
\includegraphics[width = \linewidth]{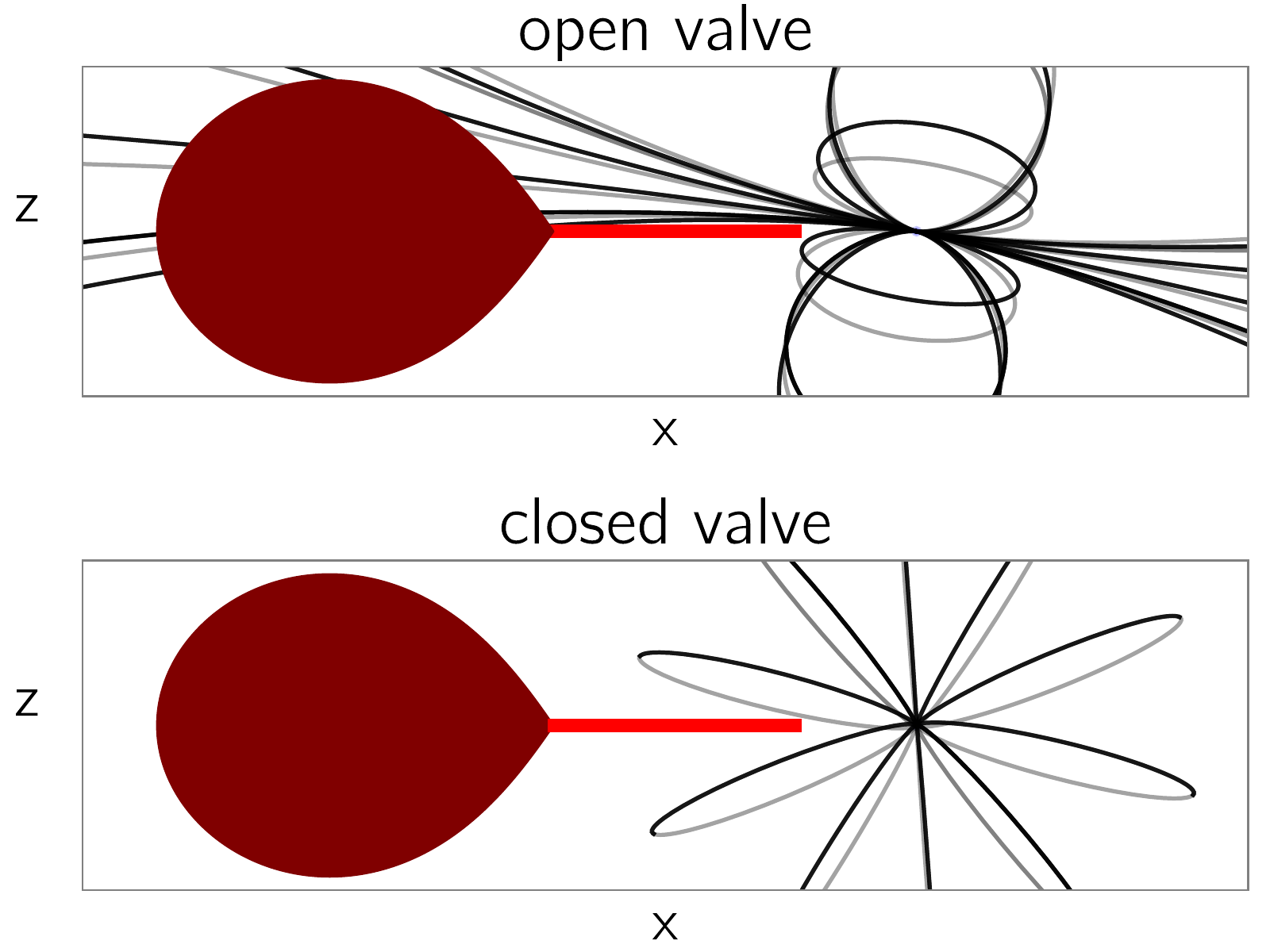}
\includegraphics[width = \linewidth]{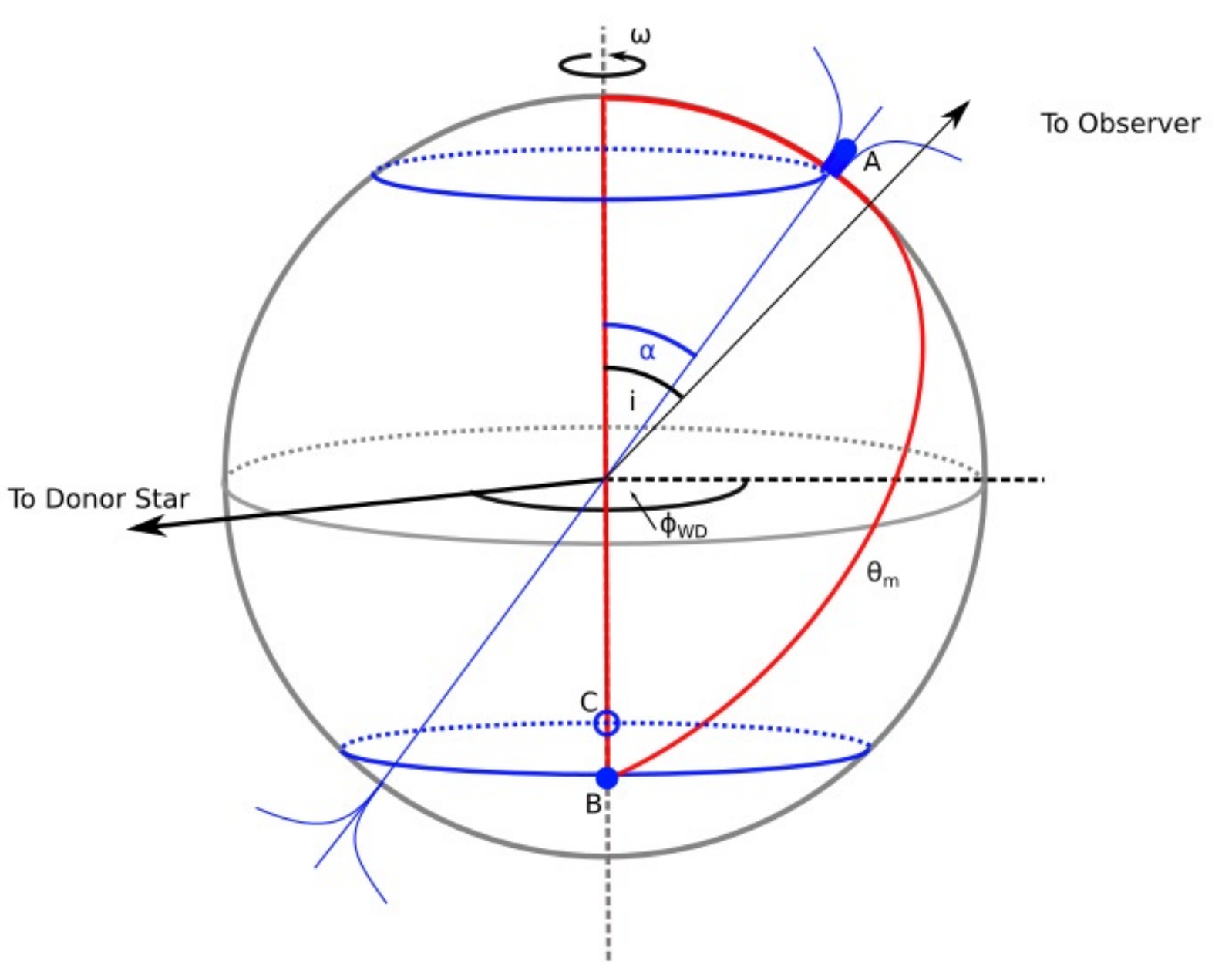}
 \figcaption{{\bf Top and middle:} Schematic diagram showing two magnetic field orientations with respect to the Roche-lobe filling donor star. The x-axis is defined by the stellar line-of-centers and the z-axis is perpendicular to the orbital plane. The red line shows the ballistic portion of the stream. Here, we view the system from within the orbital plane at orbital phase 0.75, and the two snapshots are separated by 0.25 beat cycles. The WD magnetic field is a barrier to mass transfer through L1 when the flow is perpendicular ({\bf middle}) to the magnetic field according to the equation of motion, Equation 3b. {\bf Bottom:} Schematic diagram showing the derived inclination = 43$^o$ and the best-effort determination of spot positions (see text). Accretion region A, at the upper magnetic pole, remains in view as the WD spins. Regions B and C are self-eclipsing and alternate in activity around the beat cycle. These spot positions compare favorably with MHD calculations \citep{Zhilkin12,Zhilkin16}.
\label{fig:valve}}
\end{figure}

 In the following best-effort spot analysis, in addition to a polar, non-eclipsed, accretion column, labeled A, there are two alternating accretion regions, B and C, see the bottom panel of Figure \ref{fig:valve}. They are both self-eclipsing regions located at approximately the same magnetic co-latitude of 124$^o$ (34 degrees below the magnetic equator) and separated by $\sim$ 180$^o$ in phase. The solution depends on the particular selection of the polar spot, so it is not unique. 

Despite the variable accretion geometry and other complications resulting in the phase-dependent drifting seen in Figure \ref{fig:Phase_Vs_Time_max}, several constraints on the accretion spot positions may be derived. The circular polarization is often positive then it suddenly drops to 0 or negative values, \citep{Mason89, Piirola94}, when another accretion region comes into view as the WD rotates. This suggests the existence of a self-eclipsing pole along with one that remains in view at all times. There is a stand-still and sometimes a dip in the circular polarization before the sudden drop. This is interpreted as a cyclotron beaming effect which occurs when the observer is viewing the cyclotron emitting column most directly. This cyclotron dip occurs only when the magnetic axis is within $\sim$ 10$^o$ of the line of sight. Finally, there is no linear polarization pulse, due to cyclotron beaming, which occurs when a cyclotron emitting region is perpendicular to the line of sight. If it is assumed that the WD magnetic axis is aligned with the positively polarized spot, then the magnetic co-latitude is $\alpha \sim$ 38$^o$, which we adopt. Hence, the polar spot, labeled A in Figure \ref{fig:valve}: bottom, remains in view of the observer at all times. 

We are able to use the self-occultation criterion of \citet{Chanmugam78} to constrain pole location(s) namely $i+\alpha < \ \pi/2 - \delta_s $, where $\alpha$ is the magnetic co-latitude and $\delta_s$ is the spot size. The position of self-eclipsed spots may be derived from the duration of the self eclipse, which is given by

$$ {\rm \Delta \phi = \frac{1}{\pi}cos^{-1}\big({\cot \alpha_s \cot i - \frac{\sin \delta_s}{\sin \alpha_s \sin i}\big)}, \ \ \     (1)}$$
\noindent
where $\alpha_s$ is the magnetic co-latitude of the spot \citep{Chanmugam78}. Durations vary during the beat cycle, but both self-eclipsing spots have occultation of about $\Delta \phi$ = 0.60, which from the self-eclipse criterion and Equation 1, yields a rotational co-latitude of $\alpha_s$ = 135$^0$ +/- 10$^o$, for both spots, which is 45 degrees below the orbital plane. The uncertainty is mostly based on a range of potential spot sizes $\delta_s$ = 5$^o$ - 30$^o$. Variations in the cyclotron column height are probably responsible for the beat-phase dependent self-eclipse duration. The phase difference between spots A and B is $\psi$ = 0.25. These circumstances allow the solution of the spherical triangle shown (in red) in the schematic diagram of Figure \ref{fig:valve} (bottom). The polar, non-eclipsed, accretion column is labeled A. Two additional accretion regions, B and C, are both self-eclipsing and located at approximately the same magnetic co-latitude of 124$^o$ (34 degrees below the magnetic equator) and separated by $\sim$ 180$^o$ in phase. These non-dipolar, but oppositely positioned regions, alternate in activity around the spin-orbit beat cycle. These spot positions compare favorably with MHD calculations \citep{Zhilkin12,Zhilkin16}, for an assumed dipole plus quadrupole magnetic field.

\begin{figure*}
  \centering
  \includegraphics[width= 7.0 in]{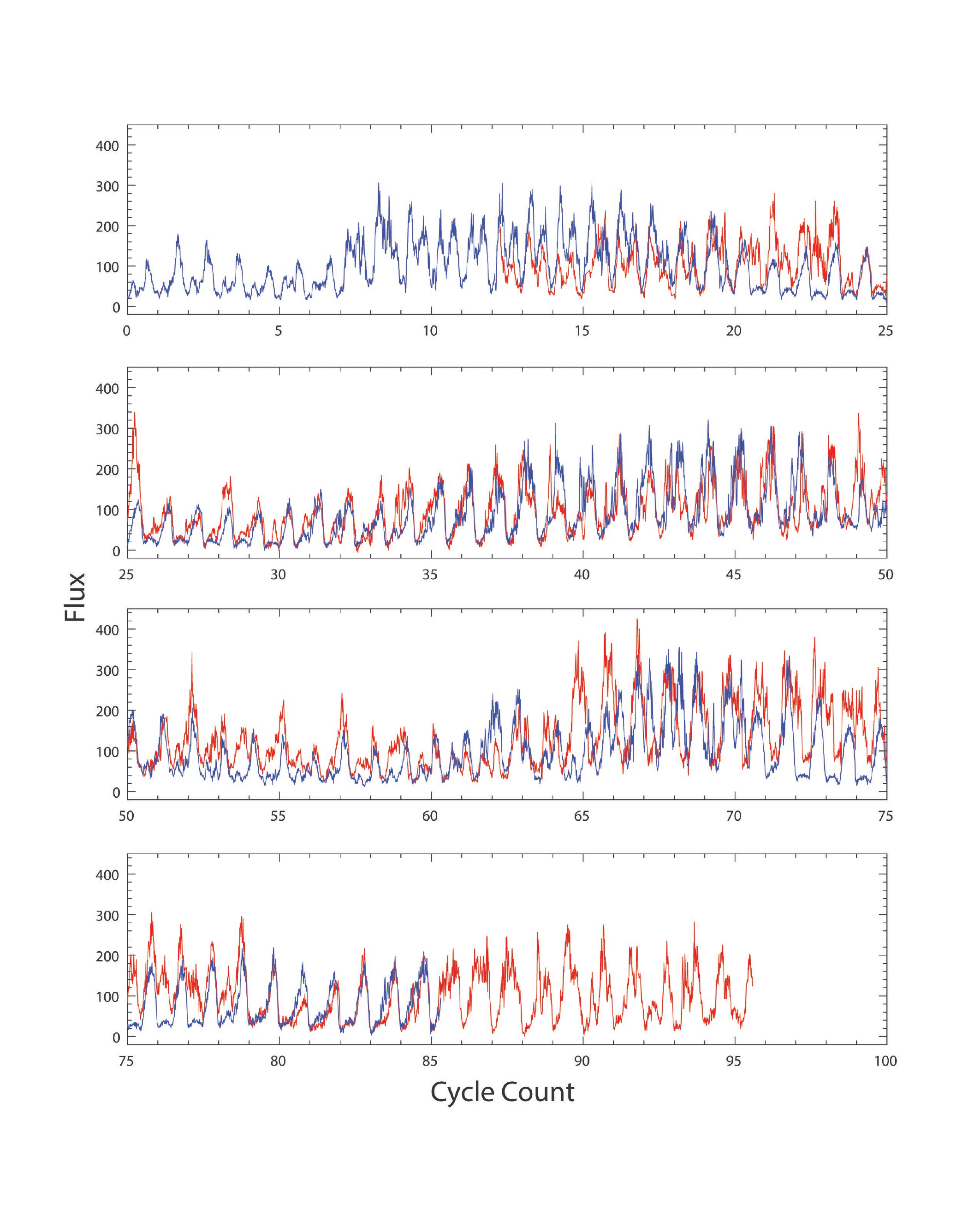}
\vskip -.6in
  \figcaption{TESS photometry of BY Cam with part 1 shown in red and part 2 again shown in blue. WD Spin cycle count 0 corresponds to the start of blue. Here they are overlaid assuming a beat period of 14.4-days, set to be exactly 104 white dwarf spin cycles. About 70$\%$ of the spin-orbit beat cycle is covered by both observations.  Notice how the structure of the blue and red curves are often very similar. Specifically, locations of photometric maxima repeat from one cycle to the next. However, peak intensities are seen to vary dramatically. The largest departures between the two curves are near cycles 25 and 65 where there is enhanced accretion in the red curves.
  \label{fig:beat}}
\end{figure*}

However, the factor of 5, beat-phase dependent, variations in the orbital averaged light curve are not easily explained by pole switching even from a hidden pole. The interpretation that accretion takes place onto a pole that is permanently in view suggests that a permanently hidden pole, oppositely opposed to pole A (Figure \ref{fig:valve}) might accrete during part of the beat cycle. However, switching from and to this region is expected to occur at most once per beat cycle. The light curves near cycle 7 in Figure \ref{fig:beat} might be evidence for a transition from a hidden pole. However, the cross-correlation, Figure \ref{fig:CC}, shows a correlation at 7.26 days, which is half of the beat cycle, and weak or no correlation at the beat period. The hidden pole model suggests that the measured brightness will modulate at the beat period, as was found in CD Ind \citep{Mason20}. The orbit-averaged light curve, top of Figure \ref{fig:2D}, shows sharp variations several times during the beat cycle, which requires some additional explanation. Beaming effects change the measured flux at a constant accretion rate, when pole switching takes place, however the intensity increase from cycles 30 to 40 and the increase from 60 to 70 in Figure \ref{fig:beat} do not involve changing of a pole position or the appearance of a new pole, rather a gradual increase in the accretion rate onto one or both active visible poles. Short-term changes in the mass transfer rate through L1 is one possibility. These variations are largely repeatable from one beat cycle to the next, Figure \ref{fig:beat}, so it is reasonable to say that if the variability is due to mass transfer rate changes, then the beat-phase dependent orientation of the WD magnetic field might be responsible. The orientation of the field is critical in this model and the L1 point crosses the WD magnetic equator twice per beat cycle, providing an explanation for the 7.26 day periodic signal. So, while it is difficult to disentangle the effects of a hidden pole and those that might depend on the changing WD magnetic field orientation, we offer a magnetic valve model as a potential explanation of the beat phase dependent intensity variations.

\section{Magnetic valve} 
\label{sec:valve}

Asynchronous polars (APs) may be thought of as a normal polar with a periodic variation of the WD magnetic field, which is assumed not to be aligned with the WD spin. So the magnetic field is variable with respect to the donor star and the accretion stream. This slow variation around the relatively long spin-orbit beat cycle of APs is in contrast with the IPs, where the WD typically spins a factor of 10 faster than the binary orbit. APs are also different than IPs, in this context, because the latter (usually) have a disk, which screens the weaker WD magnetic field from the inner Lagrangian (L1) point. Since the spin and orbital periods are identical in synchronized polars, the magnetic field structure is locked in the frame of the binary. Hence, in synchronous polars, only variations in the mass-transfer rate will result in changes in the position of the stream's magnetic threading region. 

The WD magnetosphere defines the location of the threading region, which is found by equating the ram pressure of the gas with the magnetic pressure, 
$$ \frac{B^2}{8 \pi} = \rho v^2 \ \ \ \ \ \ \ \       (2) $$
where $B$ is the magnetic field strength, $ \rho$ is the density, and $v$ is the velocity of the stream. Far from the surface of the WD, the magnetic field is approximately dipolar because higher order terms fall off more rapidly with distance from the WD.

At a given distance r, from the WD, the field above the poles is a factor of two stronger than it is at the magnetic equator. That means that as an AP progresses through its spin-orbit beat cycle $B_{L1}$ will vary between a maximum value, when the magnetic axis points most directly towards the donor (Figure \ref{fig:valve}: top) increasing the natural magnetic barrier due to the donor field, to a minimum value when L1 crosses the WD magnetic equator  (Figure \ref{fig:valve}: middle), where the WD magnetic field $B_{L1}$ periodically weakens.  

The WD magnetic moment may be written as:
\newcommand{\uvec}[1]{\boldsymbol{\hat{\textbf{#1}}}}
$$ \Vec{m} = \frac{1}{2} B_p R^3 \Vec{m}(t)\ \ \ \       (3) $$
where $\Vec{m}(t)$ is the time dependent part of the WD magnetic moment far from the WD, see \citet{Shapiro83} equation (10.5.3), $B_p$ is the measured polar magnetic field strength, R is radius of the WD, and

$$  \Vec{m}(t) =  \uvec{e}_{1}\cos\alpha + \uvec{e}_{2}\sin\alpha \cos\Omega_b t + \uvec{e}_{3}\sin\alpha \sin\Omega_b t $$

\noindent
where $\alpha$ is the magnetic co-latitude, which is the angle between the WD spin axis and the magnetic moment, $\Omega_b = \omega -\Omega$, is the spin-orbit beat frequency, and t is time. Unit vectors $\uvec{e}_{1}, \uvec{e}_{2}, \uvec{e}_{3} $ are in the direction parallel to the rotation axis and in two mutually orthogonal directions in the orbital plane, respectively. Without loss of generality, we set $\uvec{e}_{2} \perp \Vec{v}$, and $\uvec{e}_{3} \parallel \Vec{v}$ where $\Vec{v}$ is the stream velocity through L1. Dynamo theory suggests that the field increases with the stellar rotation rate and is approximately aligned with the rotational axis, so that $\Vec{B_2} \parallel \uvec{e}_{1}$.

The combined donor and WD magnetic fields at L1 result in a time-dependent field opposing the flow. The field in the $\uvec{e}_{1}$ direction is time independent. Unit vector $\uvec{e}_{3}$ is time variable, but is parallel to and hence does not oppose the flow. On the other hand, the strength of the component of $\Vec{B}$ in the $\uvec{e}_{2}$ direction changes as a function of beat phase, going to zero as L1 crosses the magnetic equator (Figure \ref{fig:valve}: middle) twice per beat cycle, at $\Omega_b t$ = 90$^o$ and 270$^o$. 

Consider the equation of motion of material at L1:
$$ \rho \frac{dv_\parallel}{dt} = f_{\parallel} \ \ \ \ \ \ \ \ \ (3a) $$
$$ \rho \frac{dv_\perp}{dt} = f_\perp - \frac{\sigma B^2} {c^2} \Delta v_\perp, \ \ \ (3b) $$
where $\rho$ is the fluid density, $f_\parallel$ and $f_{\perp}$ are the sum of the non-electromagnetic forces parallel and perpendicular to the magnetic field acting to accelerate material through L1, and $\sigma$ is the fluid conductivity. The difference between the perpendicular velocity component and the drift speed of the individual particle guiding centers is $\Delta v_{\perp}$, under the influence of electric and magnetic fields, see the discussion by \citet{Meintjes04} in the context of the IP, AE Aqr.  Figure \ref{fig:valve} shows the range of possible orientations of the magnetic field with respect to the donor. The second term of Equation 3b, shows that motion across the magnetic field is impeded. So, stronger non-electromagnetic forces, f, are needed to cross L1 in a direction that is perpendicular to the magnetic field. When the L1 point is at the magnetic equator the WD magnetic field $B$ is directed perpendicular to the flow at L1. The stream leaving L1 has a dense core with something like a Gaussian distribution from the center. In other words, when the flow through L1 experiences a magnetic field perpendicular to the flow, the magnetic valve closes and only the densest part of the flow, namely the core, is permitted to pass through L1. The lower density part of the stream is stopped. Note that since BY Cam experiences low-states vary rarely \citep{Szkody90}, the magnetic valve is capable of restricting, but not usually eliminating, the mass-transfer rate at L1 in BY Cam.

The strength of the WD magnetic field at L1 may be calculated from the stellar masses, M$_1$ = 0.76 M$_\odot$ \citep{Shaw20}, M$_2$ = 0.2 M$_\odot$ {\citep{Knigge11}}, Kepler's Third Law, the relationship between the Roche-lobe radius and the mass ratio \citep{Warner95}, along with the surface magnetic field strength. Measuring the magnetic field strength using cyclotron harmonics is especially difficult when more than one pole is active. \citet{Cropper89} derived a magnetic field strength of 40.8 MG from cyclotron humps in the optical/IR spectrum.  \citet{Tutar17} obtained spectropolarimetry at two orbital phases, a few hours apart, when the circular polarization had changed sign, see also \citet{Mason89}. They derived a magnetic field strength of 168 MG for one of the poles, while a field strength of 70, 160, or 212 MG is possible for the other pole. The \citet{Tutar17} result suffers because only one very broad cyclotron hump is seen in each of their spectra as they do not extend as far to the red as those of \citet{Cropper89}. Usually 2 or more well defined cyclotron features are considered reliable, and broad features are difficult to interpret \citep{Wickramasinghe91}. The difference might be reconciled if two sets of cyclotron features are in the unpolarized spectrum of \citet{Cropper89}. In any case, the WD field appears to be strong enough to consider the possibility that it has an effect on the flow at L1. With that in mind, using B$_{WD}$ = 168 MG and assuming a dipolar structure yields a maximum field strength at L1 of B$_{L1}$ = 0.44 kG, and if B$_{WD}$ = 40.8 MG then the maximum WD field at L1 is B$_{L1}$ = 106 G. It is the non-zero magnetic co-latitude and asynchronism which results in the variable field at L1. The field is strongest near the magnetic axis  (Figure \ref{fig:valve}: top)  and a factor of 2 lower when L1 crosses the magnetic equator (Figure \ref{fig:valve}: middle), twice per beat cycle.  Using the derived WD magnetic co-latitude of $\alpha$ = $38^o$, yields an oscillating field strength of B$_{L1}$ = 0.30 - 0.44 kG, using B$_{WD}$ = 168 MG and if B$_{WD}$ = 40.8 MG, then B$_{L1}$ = 73 - 106 G. Note that although the local field near the WD might be complex, the far-field (at L1) is dominated by the dipolar component.

We also have to consider the magnetic field of the donor at L1, which is added vectorially to the WD field. If the combination provides significant magnetic pressure perpendicular to the flow, then opening and closing of the valve will occur. The donor field will act as a barrier against the flow of matter across L1, so the ram pressure at L1 even in non-magnetic CVs must also breech the magnetic barrier of its donor. Radio emission of mCVs, from cyclotron masers near the donor reveal donor fields of a few thousand Gauss \citep{Barrett17, Barrett20}. Globally averaged magnetic fields of about 1-5 kG have also been measured in several fast rotating isolated M-dwarfs \citep{Kochukhov21}. The donor field, assumed to be dipolar and aligned with its spin, is weakest and perpendicular to the surface at L1. Hence, the strongest B-field occurs when the magnetic valve is open (Figure \ref{fig:valve}: top). The WD magnetic field at L1, modulating between 0.30 - 0.44 kG, likely does not dominate the total magnetic field strength at L1, yet the magnetic pressure variation in the $\uvec{e}_{2}$ direction has a negligible contribution from the donor, and thus is dominated by the time variable WD magnetic field. The beat-phase dependent mass transfer variations are the result of the operation of the magnetic valve. This has some theoretical basis given that \citet{Mukai88} proposed and \citet{Cash02} concurred that some portion of accreted matter is threaded onto the field at L1 in many polars. Changes in the mass-transfer rate may also be reflected in the magnetic threading distance.

The magnetic threading distance, $D_{th}$, as measured from the center of the WD, may be derived from the balance of ram pressure against magnetic pressure (Equation 1) and following \citet{Mukai88}, 
$$ D_{th} \propto {B_1^{4/11} \ R^{12/11}_1 \over M_1^{1/11}\ \dot M^{2/11}}(1+3 \rm \ cos^{2} \theta_m)^{1/6}, \ \       (4) $$
$$ D_{th} \propto B_1^{4/11} R^{37/33}_1 \dot M^{-2/11} (1+3 \rm \ cos^{2} \theta_m)^{1/6}, \ \       (5) $$
where $B_1$ is the WD surface field, $R_1$ and $M_1$ are the mass and radius of the WD respectively, {$ \dot M$} is the mass transfer rate, and here we have included the angular dependence on the magnetic latitude of the threading region and a WD mass-radius radiation $M \propto R^{1/3}$. Equation 4 should be applied with caution as the threading region is complex and specific details are important, so here it is used only to illustrate the main dependencies. Note that $\theta_m$ remains constant with respect to the stream in synchronous polars, so that only the weak mass-transfer rate dependence affects $D_{th}$. However for APs, changes in the magnetic latitude of the threading region becomes the dominant affect in determining the threading distance, see the discussion of variable threading distance in the context of the AP, V1432 Aql \citep{Littlefield15}. But we are not done yet, because as we we see in Figure \ref{fig:2D}, the accretion rate is modulated by the magnetic valve, by a factor of 5, over the beat cycle. From Equation 4, a factor of 5 increase in {$ \dot M$} will bring the threading distance 25\% closer to the WD. When the WD magnetic field orientation is advantageous for mass transfer through L1, the light curve is bright and accretion onto two poles is likely. The beat-phase dependence of the accretion rate, {$ \dot M$}, combined with the variability in $\theta_m$, would mean up to a 60$\%$ variation in $D_{th}$ over the course of the beat cycle in this case. In particular, the magnetic valve model predicts, that during a period of increased mass transfer due to the opening of the valve, there will be drifting in photometric maxima from the WD spin period, forward in phase. This effect is observed and discussed in the next section.

\section{Discussion} \label{sec:discussion}
  
In order to investigate beat-phase dependent light curve variability, consecutive beat cycles are overlaid in Figure \ref{fig:beat}. Part 1 is again shown in red and part 2 is in blue. The data were first phased with the spin period of the white dwarf P$_{S}$ = 199.384 min, then 104 cycles were subtracted from part 2 of the data (blue), in order to produce the overlay light curve shown in Figure \ref{fig:beat}. 

The light curves undergo several significant changes over the course of the beat-cycle. One thing that can have a dramatic effect on the light curve of an AP is the possibility of active accretion onto a hidden accretion spot. Non-eclipsing CVs have a portion of the WD surface permanently hidden from the observers view. The transition from low mean flux to dramatically higher flux at cycle 7, of Figure \ref{fig:beat}, is consistent with pole switching from a hidden pole to a visible two-pole configuration by cycle 10. This two-pole configuration, cycles 10 - 15, which is explained by two spots on the same hemisphere, with one spot that is always in view and the other is self-eclipsed. Pole switching to and from a hidden pole is expected to occur at most once over the course of the beat cycle. 

Comparing the red and blue light curves of Figure \ref{fig:beat} reveals both well-matched beat phase behavior as well as occasional large differences. For example, notice the significant differences seen at cycle count 25 and 65. A short-term accretion rate variation, plausibly due to magnetic activity near L1, in the red curve is the likely cause of these beat phases anomalies, and the temporary penetration of the magnetic barrier is a likely cause. In our interpretation, the magnetic valve slowly opens from cycles 30 to 40 during a broad single-pole configuration. Low minima during this time suggest that the permanently in view pole is not very active at this time. A complicated and bright blended configuration is seen around cycles 40 to 45. At low mean intensity, around cycles 55 to 60, low-level accretion onto a second pole is seen. The two-pole flaring state configuration returns during cycles 65 to 75.

The spot positions are most directly explained by the presence of a complex magnetic field in BY Cam \citep{Piirola94, Wu96, Mason96, Zhilkin12}. In particular, we note that the threading distance should be a determining factor in the nature of the magnetic accretion geometry of polars, since higher order multipolar terms weaken with increasing distance, leaving a pure dipole field at sufficiently large distances. Our best-effort solution to accretion spot location yields three regions, one is a dipolar column that does not self eclipse, while the other two are self-eclipsing and alternate in activity around the beat cycle. The presence of a complex field has implications for the magnetic valve model as well, because the measured magnetic field strengths \citep{Cropper89, Tutar17} may or may not represent the dipolar field. If the dominate cyclotron source originated from a spot near the magnetic equator, then the magnetic moment of the WD would be a factor of 2 higher. The observed complex emission lines are consistent with multiple streams, also pointing to significant magnetic influence by the WD field at L1. Our results compare favorably to those of \citet{Piirola94}, where UBVRI polarimetry was obtained and  the cyclotron emission was modelled. They fortuitously assumed i = 45$^o$, which compares well with the new dynamical value of i = 43$^o$, making the comparison with their modelling relatively easy, despite the use of an incorrect spin period. The most interesting point is that \citet{Piirola94} needed a total of three distinct accretion regions, two main accretion spots and a third weaker one, in order to fully explain the polarimetry during a single observation. With only photometry we are able to distinguish these three spots only from pole switching around the beat cycle. Magnetic threading in APs is more complex than for synchronous polars and the effect of the accretion stream on the magnetic field must also be taken into account \citep{Hameury86}. Multi-color polarimetry appears to show that all three spots may be simultaneously active. Obtaining multi-color or spectropolarimetry at several beat phases would further constrain accretion models. 

 Now we consider the potential operation of a magnetic valve in the other APs (Table \ref{table:APs}). Besides BY Cam, CD Ind is the only AP with both a WD magnetic field  B = 11 $\pm$ 2 MG \citep{Schwope97} and a precise WD mass determination, M = 0.87 (+0.04, -0.03) $M_\odot$ \citep{Dutta2022}, the latter based on the post-shock temperature derived from NuSTAR observations. A complete analysis of CD Ind requires some knowledge of its magnetic co-latitude, which provides modulation of $B_{L1}$. However, $B_{L1}$ for CD Ind may be calculated up to a factor of 2. Relative to BY Cam, several countervailing effects are at play. The orbital period of CD Ind is a bit less than 2 hr, while BY Cam has an orbital period of just longer than 3hr. However, the measured polar magnetic field strength of BY Cam is much larger than that of CD Ind (Table 1). The maximum $B_{L1}$ for CD Ind WD is 0.11 kG, so it is possible that given a favorable magnetic co-latitude, a valve operates in CD Ind. However, donor comparison is difficult. The standard model of CV evolution argues that the donor's magnetic dynamo ceases at about an orbital period of 3 hours, when the star becomes fully convective. CVs then become disconnected, losing angular momentum only by gravitational radiation until the orbital period decreases sufficiently to fill it's Roche-lobe again, resulting in the period gap. However, \citet{Knigge11} (subsection 8.5) show that the angular momentum loss below 3 hours is several times larger than that due to gravitational radiation. So it appears that there must be an additional angular momentum loss mechanism. The most likely explanation is a weaker stellar wind and hence a magnetic field. Magnetic fields in coronal maser emission regions of the donor have been observed in both short and long period polars \citep{Barrett20}. The magnetic valve mechanism is not required to explain CD Ind, since the beat-phase dependent modulation observed in CD Ind repeats on the beat cycle, which is most easily explained by pole switching onto an accretion region that, due to the binary inclination, remains hidden from view \citep{Mason20}. For CD Ind, a magnetic value is an alternative, but disfavored, model. Hidden pole accretion probably occurs in BY Cam during part of the beat cycle.

\section{Conclusion} \label{sec:conclusion}

TESS photometry of BY Cam provides the first continuous, nearly complete, observation around the beat cycle of this enigmatic mCV. In summary, beat-phased light curves are observed to undergo pole switching, from single poles at spin-phase 0.25 (cycle 30) to opposite single poles at spin-phase 0.75 (cycle 80), with high-intensity two-pole configurations in between. TESS data shows details of these transitions for the first time. Beat-phase dependent accretion rate variations are the immediate consequence of mass-transfer rate variations at L1. This is precisely the behavior predicted by the simple magnetic valve model discussed in the previous section, as L1 crosses the magnetic equator twice during each beat-cycle as in Figure \ref{fig:valve} (middle). The accretion rate appears to modulate on the beat period between the WD spin and binary orbit. A magnetic valve operating at L1 is the proposed mechanism for constricted accretion flow through L1. The resulting beat-phase dependent {$ \dot M$} combines with the beat-phase dependent magnetic field strength B to change the accretion stream threading location D$_{th}$ by a significant factor. Changes in D$_{th}$ are manifested in the complex accretion geometry evolution around the beat period. Large changes in the threading location result in differing magnetized flows onto the surface. These variations confound period analysis as different field lines carry the material to the WD surface as a function of beat phase. Photometric maxima shift significantly as the result of beat-phase dependent mass-transfer variations. That is why only lower-flux minima are sufficiently stable to allow accurate ephemeris determination. In addition, sporadic bursts of mass transfer are observed, suggesting episodic success of ram pressure against magnetic stresses at L1, possibly due to donor magnetic activity. The magnetic valve model predicts that as the valve opens, the resulting increased mass transfer causes the threading region to move closer to the WD, resulting in drifting of the photometric maxima backwards in phase, mimicking a shorter spin period. This effect is observed, see Figure \ref{fig:beat}, starting near spin-phase 0.25, lending strong support for a variable threading distance.

\acknowledgments

We thank the TESS Guest Investigator program for providing 2-min cadence observations of BY Cam. PAM dedicates this work to the memory of his PhD advisor George W. Collins II who's emphasis on developing physically motivated models for observational data helped inspire the current work. We thank the anonymous referee for suggestions that significantly improved the paper and we thank Paul Barrett, Peter Biermann, Pieter Meintjes, and Elena Pavlenko for discussions on this topic. 

PAM, LCM, and JFM acknowledge support from Picture Rocks Observatory. PS and CL acknowledge support from NSF grant AST-1514737. MRK acknowledges support from the Irish Research Council in the form of a Government of Ireland Postdoctoral Fellowship (GOIPD/2021/670: Invisible Monsters). PG and CL acknowledge support from NASA grants 80NSSC19K1704 and 80NSSC22K0183. GR acknowledges Armagh Observatory \& Planetarium which is core funded by the Northern Ireland Executive. 


\bibliography{bib.bib}

\end{document}